\RequirePackage{fix-cm}
\documentclass[smallextended]{svjour3}       
\smartqed  
\usepackage{graphicx}
\usepackage{amsmath,amssymb}

\usepackage{changepage}

\usepackage[utf8x]{inputenc}

\usepackage{textcomp,marvosym}

\usepackage{cite}

\usepackage{nameref,hyperref}

\usepackage[right]{lineno}

\usepackage{microtype}
\DisableLigatures[f]{encoding = *, family = * }

\usepackage[table]{xcolor}

\usepackage{array}

\begin{document}

\title{Classification of COVID-19 in chest X-ray images using DeTraC deep convolutional neural network
}

\titlerunning{Classification of COVID-19 in chest X-ray images using DeTraC}        

\author{Asmaa Abbas         \and
        Mohammed M. Abdelsamea \and
        Mohamed Medhat Gaber
}


\institute{Asmma Abbas \at
              Mathematics Department, Faculty of Science, Assiut University, Assiut, Egypt \\
              \email{asmaa.abbas@science.aun.edu.eg}           
           \and
           Mohammed M. Abdelsamea \at
              School of Computing and Digital Technology, Birmingham City University, Birmingham, UK and Mathematics Department, Faculty of Science, Assiut University, Assiut, Egypt \\
              \email{mohammed.abdelsamea@bcu.ac.uk}           
           \and
           Mohamed Medhat Gaber \at
              School of Computing and Digital Technology, Birmingham City University, Birmingham, UK\\
              \email{mohamed.gaber@bcu.ac.uk}           
           \
}

\date{Received: date / Accepted: date}

\maketitle

\begin{abstract}
Chest X-ray is the first imaging technique that plays an important role in the diagnosis of COVID-19 disease. Due to the high availability of large-scale annotated image datasets, great success has been achieved using convolutional neural networks (\emph{CNN}s) for image recognition and classification. However, due to the limited availability of annotated medical images, the classification of medical images remains the biggest challenge in medical diagnosis. Thanks to transfer learning, an effective mechanism that can provide a promising solution by transferring knowledge from generic object recognition tasks to domain-specific tasks. In this paper, we validate and adapt a deep \emph{CNN}, called Decompose, Transfer, and Compose (\emph{DeTraC}), for the classification of COVID-19 chest X-ray images. \emph{DeTraC} can deal with any irregularities in the image dataset by investigating its class boundaries using a class decomposition mechanism. The experimental results showed the capability of \emph{DeTraC} in the detection of COVID-19 cases from a comprehensive image dataset collected from several hospitals around the world. High accuracy of 95.12\% (with a sensitivity of 97.91\%, and a specificity of 91.87\%) was achieved by \emph{DeTraC} in the detection of COVID-19 X-ray images from normal, and severe acute respiratory syndrome cases.
\end{abstract}


\section{Introduction}
\begin{figure}[]
    \centering

        \includegraphics[scale=0.6]{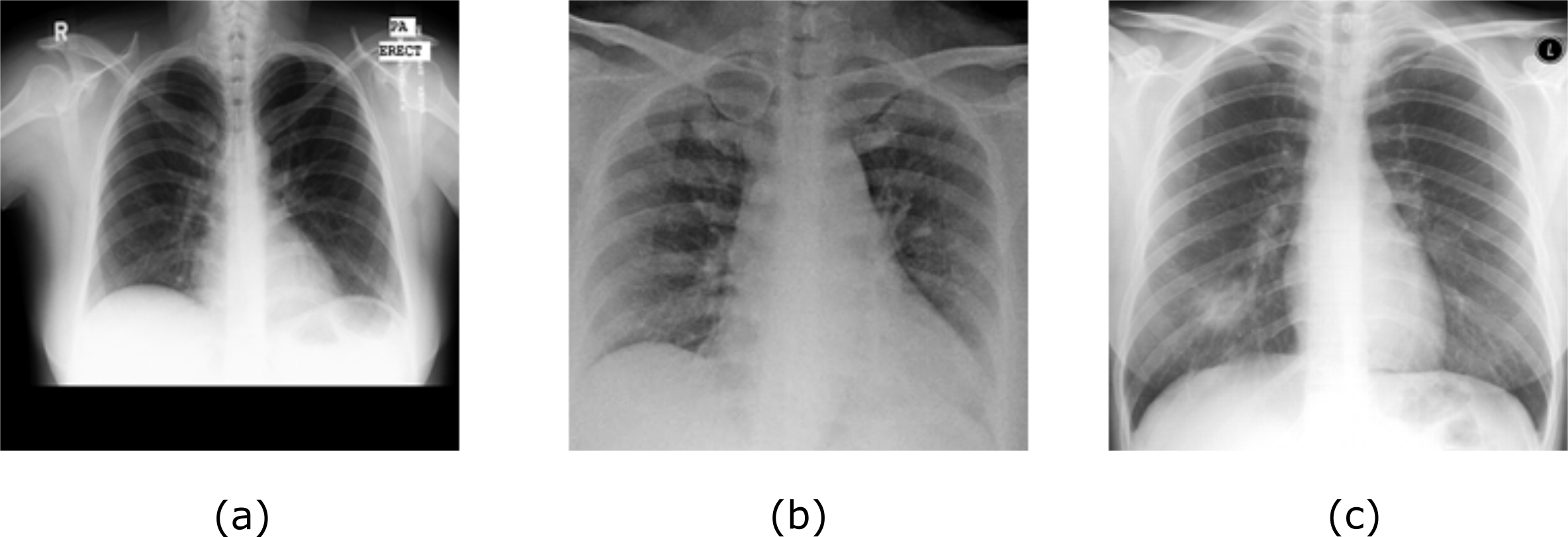}
        \caption{Examples of a) normal, b) COVID-19, and c) SARS chest x-ray images.}
        \label{Figimage}
    \end{figure}

Diagnosis of COVID-19 is typically associated with both the symptoms of pneumonia and Chest X-ray tests \cite{shi2020radiological}. Chest X-ray is the first imaging technique that plays an important role in the diagnosis of COVID-19 disease. Fig. \ref{Figimage} shows a negative example of a normal chest x-ray, a positive one with COVID-19, and a positive one with the severe acute respiratory syndrome (SARS). 

Several classical machine learning approaches have been previously used for automatic classification of digitised chest images \cite{dandil2014artificial,kuruvilla2014lung}. For instance, in \cite{manikandan2016lung}, three statistical features were calculated from lung texture to discriminate between malignant and benign lung nodules using a Support Vector Machine \emph{SVM} classifier. A grey-level co-occurrence matrix method was used with Backpropagation Network \cite{sangamithraa2016lung} to classify images from being normal or cancerous. With the availability of enough annotated images, deep learning approaches \cite{anthimopoulos2016lung, sun2016computer, abbas2018learning} have demonstrated their superiority over the classical machine learning approaches. \emph{CNN} architecture is one of the most popular deep learning approaches with superior achievements in the medical imaging domain \cite{lecun2015deep}. The primary success of \emph{CNN} is due to its ability to learn features automatically from domain-specific images, unlike the classical machine learning methods. The popular strategy for training \emph{CNN} architecture is to transfer learned knowledge from a pre-trained network that fulfilled one task into a new task \cite{pan2009survey}. This method is faster and easy to apply without the need for a huge annotated dataset for training; therefore many researchers tend to apply this strategy especially with medical imaging. Transfer learning can be accomplished with three major scenarios \cite{li2014medical}: a) ``shallow tuning'', which adapts only the last classification layer to cope with the new task, and freezes the parameters of the remaining layers without training; b) ``deep tuning'' which aims to retrain all the parameters of the pre-trained network from end-to-end manner; and (c) ``fine-tuning'' that aims to gradually train more layers by tuning the learning parameters until a significant performance boost is achieved. Transfer knowledge via fine-tuning mechanism showed outstanding performance in X-ray image classification \cite{anthimopoulos2016lung, shin2016deep, gao2018holistic}.

Class decomposition \cite{vilalta2003class} has been proposed with the aim of enhancing low variance classifiers facilitating more flexibility to their decision boundaries. It aims to the simplification of the local structure of a dataset in a way to cope with any irregularities in the data distribution. Class decomposition has been previously used in various automatic learning workbooks as a preprocessing step to improve the performance of different classification models. In the medical diagnostic domain, class decomposition has been applied to significantly enhance the classification performance of models such as Random Forests, Naive Bayes, C4.5, and $SVM$ \cite{polaka2010clustering, polaka2013clustering, wu2010cog}. 

In this paper, we adapt our previously proposed convolutional neural network architecture based on class decomposition, which we term  Decompose, Transfer, and Compose ($DeTraC$) model, to improve the performance of pre-trained models on the detection of COVID-19 cases from chest X-ray images \footnote{The developed code is available at https://github.com/asmaa4may/DeTraC\_COVId19.}. This is by adding a class decomposition layer to the pre-trained models. The class decomposition layer aims to partition each class within the image dataset into several sub-classes and then assign new labels to the new set, where each subset is treated as an independent class, then those subsets are assembled back to produce the final predictions. For the classification performance evaluation, we used images of chest x-ray collected from several hospitals and institutions. The dataset provides complicated computer vision challenging problems due to the intensity inhomogeneity in the images and irregularities in the data distribution.  

The paper is organised as follow.In Section \ref{relatedwork}, we review the state-of-the-art methods for COVID-19 detection. Section \ref{methods} discusses the main components of \emph{DeTraC} and its adaptation to the detection of COVID-19 cases. Section \ref{results} describes our experiments on several chest X-ray images collected from different hospitals. In Section \ref{discussion}, we discuss our findings. Finally, Section \ref{conclusion} concludes the work.

\section{Related work}
\label{relatedwork}

In the last few months, World Health Organization (WHO) has declared that a new virus called COVID-19 has been spread aggressively in several countries around the world \cite{world2020coronavirus}. Diagnosis of COVID-19 is typically associated with the symptoms of pneumonia, which can be revealed by genetic and imaging tests. Fast detection of the COVID-19 can be contributed to control the spread of the disease. 

Image tests can provide a fast detection of COVID-19, and consequently contribute to control the spread of the disease. Chest X-ray (CXR) and Computed Tomography (CT) are the imaging techniques that play an important role in the diagnosis of COVID-19 disease. The historical conception of image diagnostic systems has been comprehensively explored through several approaches ranging from feature engineering to feature learning. 

Convolutional neural network (CNN) is one of the most popular and effective approaches in the diagnosis of COVD-19 from digitised images. Several reviews have been carried out to highlight recent contributions to COVID-19 detection \cite{shi2020review, 9079648, li2020artificial}. For example, in \cite{wang2020deep}, a \emph{CNN} was applied based on Inception network to detect COVID-19 disease within computed tomography (\emph{CT}). In \cite{song2020deep}, a modified version of ResNet-50 pre-trained network has been provided to classify \emph{CT} images into three classes: healthy, COVID-19 and bacterial pneumonia. Chest X-ray images (\emph{CXR}) were used in \cite{sethy2020detection} by a \emph{CNN} constructed based on various ImageNet pre-trained models to extract the high level features. Those features were fed into \emph{SVM} as a machine learning classifier in order to detect the COVID-19 cases. Moreover, in \cite{wang2020covidnet}, a \emph{CNN} architecture called COVID-Net based on transfer learning was applied to classify the \emph{CXR} images into four classes: normal, bacterial infection, non-COVID and COVID-19 viral infection. In \cite{apostolopoulos2020covid}, a dataset of X-ray images from patients with pneumonia, confirmed Covid-19 disease, and normal incidents, was used to evaluate the performance of state-of-the-art convolutional neural network architectures proposed previously for medical image classification. The study suggested that transfer learning can extract significant biomarkers related to the Covid-19 disease.

Having reviewed the related work, it is evident that despite the success of deep learning in the detection of Covid-19 from CXR and CT images, data irregularities have not been explored. It is common in medical imaging in particular that datasets exhibit different types of irregularities (e.g. overlapping classes) that affect the resulting accuracy of machine learning models. Thus, this work focuses on dealing with data irregularities, as presented in the following section.

\section{DeTraC method}
\label{methods}

This section describes in sufficient details the proposed method for detecting Covid-19 from chest X-ray images. Starting with an overview of the architecture through to the different components of the method, the section discusses the workflow and formalises the method.

\subsection{DeTraC architecture overview}

\begin{figure*}[]
    \centering
        \includegraphics[width=\columnwidth,height=10cm]{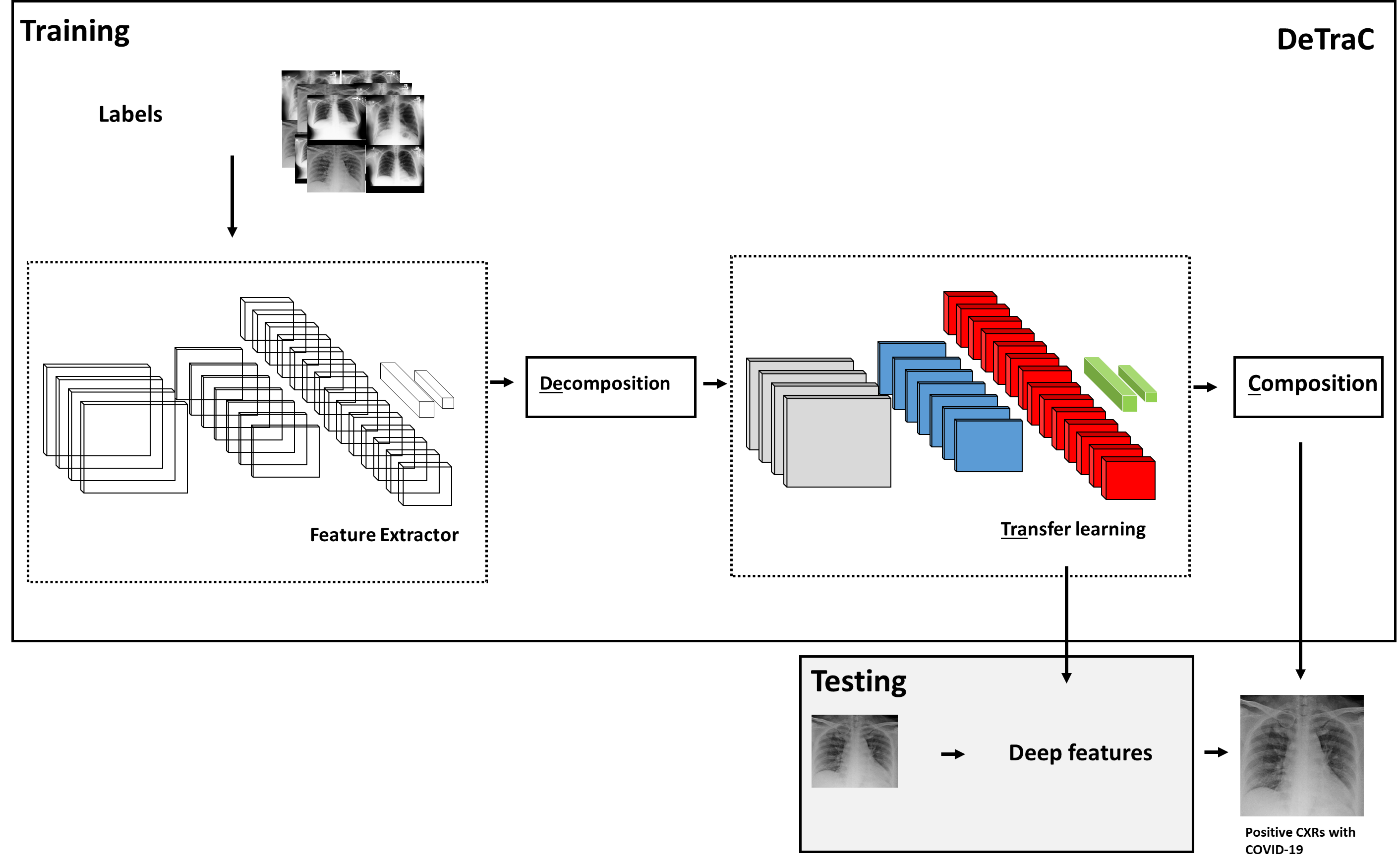}
        \caption{\underline{De}compose, \underline{Tra}nsfer, and \underline{C}ompose (\emph{DeTraC}) model for the detection of COVID-19 from chest X-ray images.}
        \label{Figproposed}
    \end{figure*}

\emph{DeTraC} model consists of three phases. In the first phase, we train the backbone pre-trained \emph{CNN} model of \emph{DeTraC} to extract deep local features from each image. Then we apply the class-decomposition layer of \emph{DeTraC} to simplify the local structure of the data distribution. In the second phase, the training is accomplished using a sophisticated gradient descent optimisation method. Finally, we use the class-composition layer of \emph{DeTraC} to refine the final classification of the images. As illustrated in Fig. \ref{Figproposed}, class decomposition and composition components are added respectively before and after knowledge transformation from an ImageNet pre-trained \emph{CNN} model. The class decomposition component aiming at partitioning each class within the image dataset into $k$ sub-classes, where each subclass is treated independently. Then those sub-classes are assembled back using the class-composition component to produce the final classification of the original image dataset.

\subsection{Deep feature extraction}

A shallow-tuning mode was used during the adaptation and training of an ImageNet pre-trained \emph{CNN} model using the collected chest X-ray image dataset. We used the off-the-shelf \emph{CNN} features of pre-trained models on ImageNet (where the training is accomplished only on the final classification layer) to construct the image feature space. However, due to the high dimensionality associated with the images, we applied \emph{PCA} to project the high-dimension feature space into a lower-dimension, where highly correlated features were ignored. This step is important for the class decomposition to produce more homogeneous classes, reduce the memory requirements, and improve the efficiency of the framework.

\subsection{Class decomposition}


		Now assume that our feature space (\emph{PCA}'s output) is represented by a 2-D matrix (denoted as dataset $A$), and \( \mathbf{L} \)  is a class category. \(  A \) and \(  \mathbf{L} \) can be rewritten as

	\begin{equation}
      A= \left[ \begin{matrix}
        a_{11}  &  a_{11}  &   \ldots  ~~~~~ a_{1m}\\
        a_{21}  &  a_{22}  &   \ldots ~~~~~~ a_{2m}\\
         \vdots   &   \vdots   &   \vdots ~~~~~~~~~~~  \vdots \\
        a_{n1}  &  a_{n2}  &   \ldots ~~~ a_{nm}\\
        \end{matrix}
         \right]  ,   \mathbf{L}= \left\{ l_{1}, l_{2}, \ldots ,l_{k} \right\},
	\end{equation}

where $n$ is the number of images,  \(  m \) is the number of features, and $k$  is the number of classes. For class decomposition, we used $k$-means clustering \cite{wu2008top} to further divide each class into homogeneous sub-classes (or clusters), where each pattern in the original class  \(  \mathbf{L}  \)  is assigned to a class label associated with the nearest centroid based on the squared euclidean distance (\emph{SED}):

	\begin{equation}
     SED= \sum _{j=1}^{k} \sum _{i=1}^{n}\parallel a_{i}^{ \left( j \right) }-c_{j}\parallel, 
	\end{equation}
	
where centroids are denoted as  \( c_{j} \).

Once the clustering is accomplished, each class in  \( \mathbf{L} \) will further divided into $k$ subclasses, resulting in a new dataset (denoted as dataset $B$). 



Accordingly, the relationship between dataset $A$ and $B$ can be mathematically described as:



	\begin{equation}
 A = ( A | \mathbf{L} ) ~ \mapsto~ B= (B | \mathbf{C} )
    \end{equation}

where the number of instances in  $A$  is equal to $B$ while $\mathbf{C}$ encodes the new labels of the subclasses (e.g. $\mathbf{C} =\{l_{11}, l_{12}, \dots, l_{1k}, l_{21}, l_{22}, \dots, l_{2k}, \dots l_{ck} \}$). Consequently $A$ and $B$ can be rewritten as:

	\begin{equation}
	    \begin{aligned}
         A= \left[ \begin{matrix}
        a_{11}  &  a_{11}  &   \ldots  ~~~~~ a_{1m }  &  l_{1}\\
        a_{21}  &  a_{22}  &   \ldots ~~~~~~ a_{2m}  &  l_{1}\\
         \vdots   &   \vdots   &   \vdots ~~~~~~~~~~~  \vdots   &   \vdots \\
         \vdots   &   \vdots   &  \begin{matrix}
         \vdots   &  ~~~~~~~  \vdots \\
        \end{matrix}
          &  l_{2}\\
        a_{n1}  &  a_{n2}  &   \ldots ~~~ a_{nm}  &  l_{2 }\\
        \end{matrix}
         \right], \\  
        B= \left[ \begin{matrix}
        b_{11}  &  b_{11}  &   \ldots  ~~~~~ b_{1m}  &  l_{11}\\
        b_{21}  &  b_{22}  &   \ldots ~~~~~~ b_{2m}  &  l_{1c}\\
         \vdots   &   \vdots   &   \vdots ~~~~~~~~~~~  \vdots   &   \vdots \\
         \vdots   &   \vdots   &  \begin{matrix}
         \vdots   &  ~~~~~~~  \vdots \\
        \end{matrix}
          &  l_{21}\\
        b_{n1}  &  b_{n2}  &   \ldots ~~~ b_{nm}  &  l_{2c}\\
        \end{matrix}
         \right].
        \end{aligned}
	\end{equation}



\subsection{Transfer learning}


With the high availability of large-scale annotated image datasets, the chance for the different classes to be well-represented is high. Therefore, the learned in-between class-boundaries are most likely to be generic enough to new samples. On the other hand, with the limited availability of annotated medical image data, especially when some classes are suffering more compared to others in terms of the size and representation, the generalisation error might increase. This is because there might be a miscalibration between the minority and majority classes. Large-scale annotated image datasets (such as ImageNet) provide effective solutions to such a challenge via transfer learning where tens of millions parameters (of \emph{CNN} architectures) are required to be trained. 

For transfer learning, we used the ImageNet pre-trained ResNet \cite{he2016deep} model, which showed excellent performance on ImageNet with only 18 layers. Here we consider freezing the weights of low-level layers and update weighs of high-level layers. 

For fine-tuning the parameters, the learning rate for all the \emph{CNN} layers was fixed to 0.0001 except for the last fully connected layer (was 0.01), the min batch size was 64 with minimum 256 epochs, 0.001 was set for the weight decay to prevent the overfitting through training the model, and the momentum value was 0.9. With the limited availability of training data, stochastic gradient descent (SGD) can heavily be fluctuating the objective/loss function and hence overfitting can occur. To improve convergence and overcome overfitting, the mini-batch of stochastic gradient descent (\emph{mSGD}) was used to minimise the objective function, $E(\cdot)$, with cross-entropy loss

\begin{eqnarray}
E\left ( y^{j},z(x^{j}) \right ) & = & -\frac{1}{n}\sum_{j=0}^{n} [ y^{j} \ln {z\left(x^{j}\right )} \nonumber\\ &&  
+ \left ( 1-y^{j} \right )\ln {\left ( 1-z\left( x^{j} \right ) \right )} ],
\end{eqnarray}

where \( x^{j} \) is the set of input images in the training, \( y^{j} \) is the ground truth labels while  \( z (\cdot)\) is the predicted output from a softmax function.    

\subsection{Evaluation and composition}


In the class decomposition layer of \emph{DeTraC}, we divide each class within the image dataset into several sub-classes, where each subclass is treated as a new independent class. In the composition phase, those sub-classes are assembled back to produce the final prediction based on the original image dataset. For performance evaluation, we adopted Accuracy (ACC), Specificity (SP) and Sensitivity (SN) metrics.They are defined as:

	\begin{eqnarray} 
	{\rm Accuracy} (ACC) & = & \frac{TP+TN}{n}\,, \\ 
	{\rm Sensitivity } (SN) & = & \frac{TP}{TP+FN}\,, 
	\\ 
	{\rm Specificity } (SP) & = & \frac{TN}{TN + FP}\,,	
	\end{eqnarray}

where $TP$ is the true positive in case of COVID-19 case and  $TN$ is the true negative in case of normal or other disease, while $FP$ and $FN$ are the incorrect model predictions for COVID-19 and other cases.

More precisely, in this work we are coping with a multi-classification problem. Consequently, our model has been evaluated using a multi-class confusion matrix of \cite{sokolova2009systematic}. Before error correction, the input image can be classified into one of ($c$) non-overlapping classes. As a consequence, the confusion matrix would be a ($ N_{c} \times N_{c} $) matrix, and $TP$, $TN$, $FP$ and $FN$ for a specific class $i$ are defined as:

\begin{eqnarray} 
    TP_{i}=\sum_{i=1}^{n}x_{ii}\\
    TN_{i}=\sum_{j=1}^{c}\sum_{k=1}^{c} x_{jk} ,j\neq i ,k\neq i\\
    FP_{i}=\sum_{j=1}^{c} x_{ji}   , j\neq i\\
    FN_{i}=\sum_{j=1}^{c} x_{ij}   , j\neq i,
\end{eqnarray}

where $x_{ii}$ is an element in the diagonal of the matrix. 
\newline
	
Having discussed and formalised the DeTraC method in this section in details, the following section validates the method experimentally. The method establishes the effectiveness of class decomposition in detecting Covid-19 from chset X-ray images. 


\section{Experimental study}
\label{results}
This section presents the dataset used in evaluating the proposed method, and discusses the experimental results. 



\subsection{Dataset}
In this work we used a combination of two datasets:
\begin{itemize}
    \item 80 samples of normal CXRs (with $4020 \times 4892$ pixels) from the Japanese Society of Radiological Technology (\emph{JSRT}) \cite{6663723, 6616679}.
    \item Chest X-ray images of \cite{cohen2020covid}, which contains 105 and 11 samples of COVID-19 and SARS (with $4248 \times 3480$ pixels).
\end{itemize}

We applied different data augmentation techniques to generate more samples such as: flipping up/down and right/left, translation and rotation using random five different angles. This process resulted in a total of 1764 samples. Also, a histogram modification technique was applied to enhance the contrast of each image.

\subsection{Class decomposition based on deep features}
We used AlexNet \cite{krizhevsky2012imagenet} pre-trained network based on shallow learning mode to extract discriminative features of the three original classes. AlexNet is composed of 5 convolutional layers to represent learned features, 3 fully connected layers for the classification task. AlexNet uses $3 \times 3$ max-pooling layers with ReLU activation functions and three different kernel filters. We adopted the last fully connected layer into three classes and initialised the weight parameters for our specific classification task. For class decomposition process, we used $k$-means clustering \cite{wu2008top}. In this step, as pointed out in \cite{Asmaa}, we selected $k$  = 2 and hence each class in  \( \mathbf{L} \)  is further divided into two clusters (or subclasses), resulting in a new dataset (denoted as dataset $B$) with six classes (norm\_1, norm\_2, COVID19\_1,COVID19\_2, SARS\_1, and SARS\_2), see Table \ref{tabsizeCXR}.

\begin{table*} 
\begin{center} 

\caption{Samples distribution in each class of chest X-ray dataset before and after class decomposition.} 
\label{tabsizeCXR} 
{
\begin{tabular}{ccccccc}
\hline
Original labels & \multicolumn{2}{c}{norm} & \multicolumn{2}{c}{COVID19} & \multicolumn{2}{c}{SARS} \\
\# instances & \multicolumn{2}{c}{80} & \multicolumn{2}{c}{105} & \multicolumn{2}{c}{11} \\
\hline
Decomposed labels & norm\_1& norm\_2  & COVID19\_1& COVID19\_2 & SARS\_1     & SARS\_2  \\
\# instances& 441  & 279 & 666  & 283  & 63  & 36\\
\hline         
\end{tabular}
} 
\end{center} 
\end{table*}

\subsection{Parameter settings and accuracy}

All the experiments in our work have been carried out in MATLAB 2019a on a PC with the following configuration: 3.70 GHz Intel(R) Core(TM) i3-6100 Duo, NVIDIA Corporation with the donation of the Quadra P5000GPU, and 8.00 GB RAM.\\

The dataset was divided into two groups; 70\% for training the model and 30\% for evaluation of the classification performance. We used ResNet18 as an ImageNet pre-trained network in the transfer learning component of \emph{DeTraC}. ResNet18 \cite{szegedy2017inception} consist of 18 layers with input image size of $224 \times 224$ and achieved an effective performance with 95.12\% of accuracy. The last fully-connected layer was changed into the new task to classify six classes.  The learning rate for all the \emph{CNN} layers was fixed to 0.0001 except for the last fully connected layer (was 0.01) to accelerate the learning. The min batch size was 64 with a minimum 100 epochs, 0.0001 was set for the weight decay to prevent the overfitting through training the model, and the momentum value was 0.95. The schedule of drop learning rate was set to 0.95 every 5 epochs. \emph{DeTraC-ResNet18} was trained based on deep learning mode. For performance evaluation, we adopted some metrics from the confusion matrix such as accuracy, sensitivity, specificity, and precision. The results were reported and summarised in table \ref{performResNet}.

\begin{table}
\large
\begin{center} 
\caption{COVID-19 classification obtained by \emph{DeTraC-ResNet18} on chest X-ray images.} 

\label{performResNet} 
\begin{tabular}{ccc}
\hline
\multicolumn{3}{c}{DeTraC- ResNet18}  \\
Accuracy     & Sensitivity      & Specificity       \\
\hline
95.12\% & 97.91\% & 91.87\%  \\
\hline
\end{tabular}
\end{center} 
\end{table}

\begin{figure}[]
\centering
\includegraphics[scale=0.38]{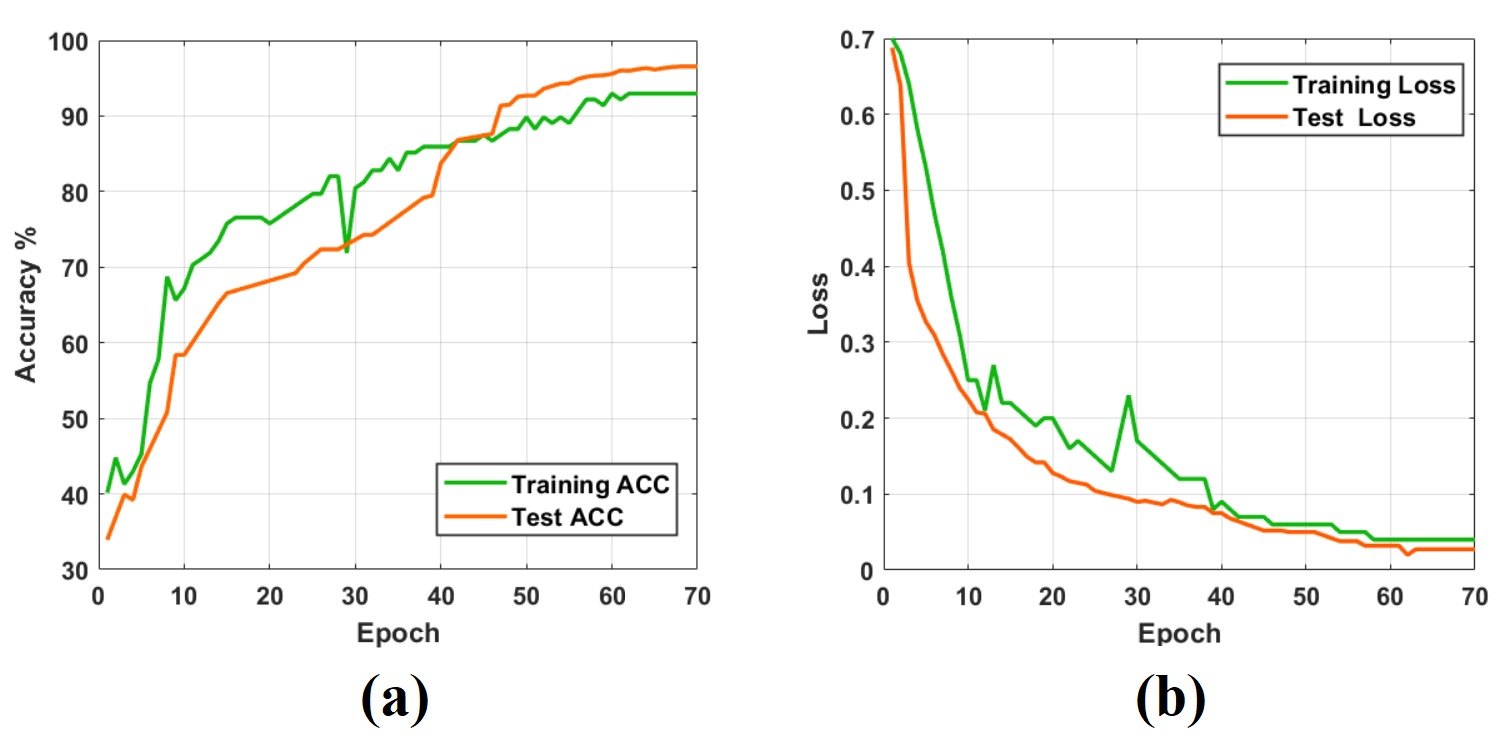}
\caption {The learning curve accuracy and error obtained by  \emph{DeTraC-ResNet18} model.}
\label{learningcurve}
\end{figure}

\begin{figure}[]
\centering
\includegraphics[scale=0.5]{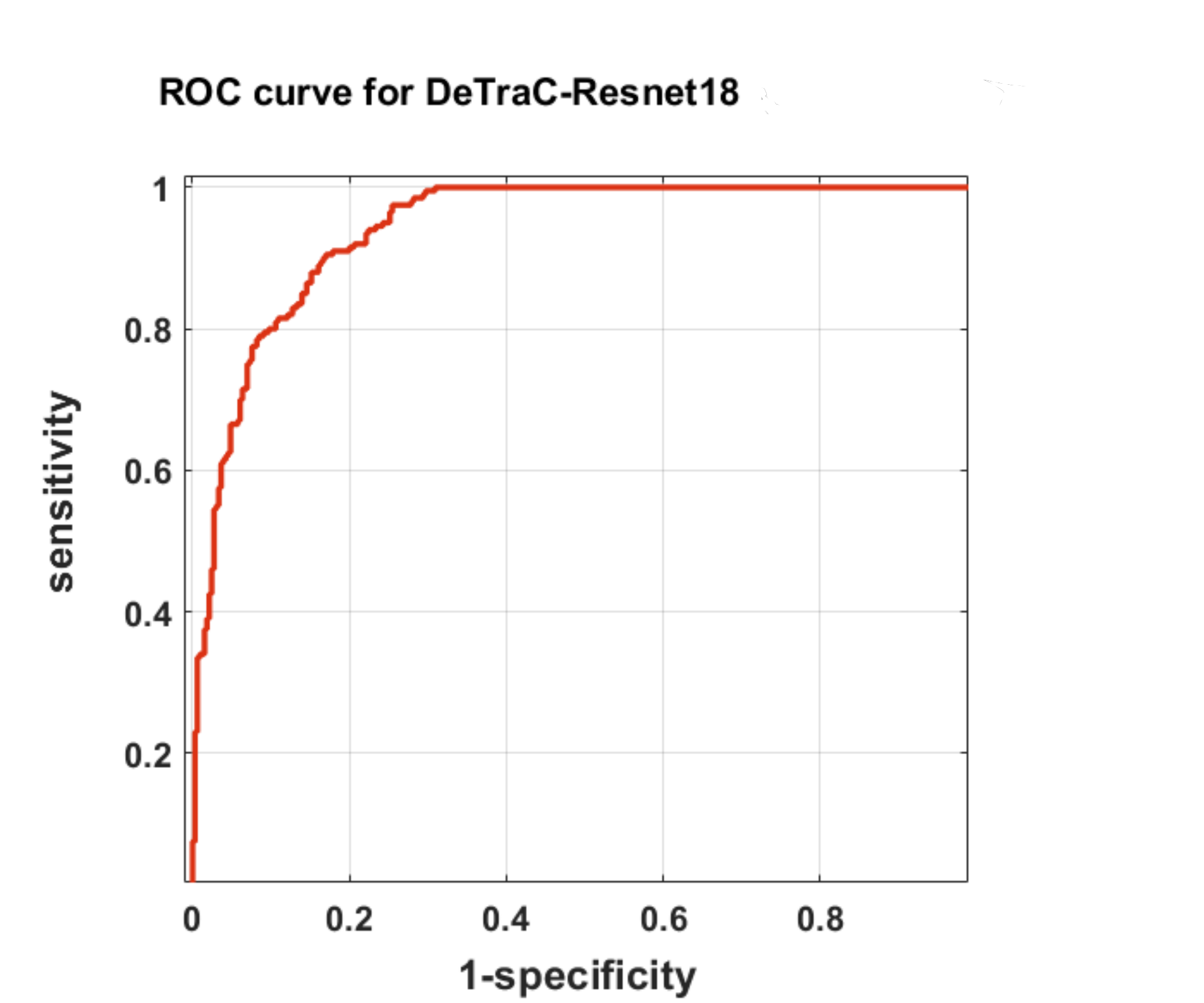}
\caption {The ROC analysis curve by training \emph{DeTraC} model based on ResNet pre-trained network}
\label{ROC}
\end{figure}

We plot the learning curve accuracy and loss between training and test as shown in Fig \ref{learningcurve}. Also, the Area Under the receiver curve (AUC) was computed as shown in Fig \ref{ROC}.

To demonstrate the robustness of \emph{DeTraC-ResNet18} in the classification of COVID-19 images, we compare it with \emph{ResNet18} using the same settings. \emph{ResNet18} achieved accuracy of 92.5\%, sensitivity of 65.01\%, and specificity of 94.3\%.

\section{Discussion}
\label{discussion}
Training \emph{CNN}s can be accomplished using two different strategies. They can be used as an end-to-end network, where an enormous number of annotated images must be provided (which is impractical in medical imaging). Alternatively, transfer learning usually provides an effective solution with the limited availability of annotated images by transferring knowledge from pre-trained \emph{CNN}s (that have been learned from a bench-marked large-scale image dataset) to the specific medical imaging task. Transfer learning can be further accomplished by three main scenarios: shallow-tuning, fine-tuning, or deep-tuning. However, data irregularities, especially in medical imaging applications, remain a challenging problem that usually results in miscalibration between the different classes in the dataset. \emph{CNN}s can provide an effective and robust solution for the detection of the COVID-19 cases from chest X-ray \emph{CXR} images and this can be contributed to control the spread of the disease. 

Here, we adopt and validate our previously developed deep convolutional neural network, we called \emph{DeTraC}, to deal with such a challenging problem by exploiting the advantages of class decomposition within the $CNN$s for image classification. \emph{DeTraC} achieved high accuracy of 95.12\% with ResNet on \emph{CXR} images. \emph{DeTraC} has demonstrated its robustness in coping with the limited availability of training images and irregularities in the data distribution. More importantly, the proposed class decomposition layer provides a generic solution to improve the efficiency of a convolutional neural network.


\section{Conclusion and future work}
\label{conclusion}

Diagnosis of COVID-19 is typically associated with the symptoms of pneumonia, which can be revealed by genetic and imaging tests. Imagine test can provide a fast detection of the COVID-19 and consequently contribute to control the spread of the disease. Chest X-ray (CXR) and Computed Tomography (CT) are the imaging techniques that play an important role in the diagnosis of COVID-19 disease.
Paramount progress has been made in deep convolutional neural networks ($CNN$s) for medical image classification, due to the availability of large-scale annotated image datasets. $CNN$s enable learning highly representative and hierarchical local image features directly from data. However, the irregularities in annotated data remains the biggest challenge in coping with COVID-19 cases from Chest X-ray images.

In this paper, we adapted \emph{DeTraC} deep \emph{CNN} architecture that relies on a class decomposition approach for the classification of COVID-19 images in a comprehensive dataset of chest X-ray images. DeTraC showed effective and robust solutions for the classification of COVID-19 cases and its ability to cope with data irregularity and the limited number of training images too. 

With the continuous collection of data, we aim in the future to extend the experimental work validating the method with larger datasets. We also aim to add an explainability component to enhance the usability of the model. Finally, to increase the efficiency and allow deployment on handheld devices, model pruning and quantisation will be utilised.

%
%
%
    \bibliographystyle{spmpsci}      
	\bibliography{references}


\end{document}